# Tunable Resonance and Electron-Phonon Coupling in Layered MoS$_2$

Deepu Kumar[1#], Nasaru Khan[1], Rahul Kumar[2], Mahesh Kumar[2], Pradeep Kumar[1*]

[1]*School of Physical Sciences, Indian Institute of Technology Mandi, 175005, India*
[2]*Department of Electrical Engineering, Indian Institute of Technology Jodhpur, 342037, India*

**Abstract**

Resonance Raman scattering, a very effective and sensitive technique for atomically thin semiconducting transition metal dichalcogenide, can be used to observe the phonons from the entire Brillouin zone. In addition to the significance of resonance effect on the Raman spectrum it may also be used to probe the electron-phonon coupling. Our study is devoted to understand the phonons in layered MoS$_2$, especially for very low frequency range ($\leq$ 100 cm$^{-1}$), as a function of temperature under the resonance effect. Understanding the phonon-phonon and electron-phonon coupling and the effects of temperature on the Raman spectrum are the central points of the present study. We observe the anomalous softening and broadening of a very low frequency phonon mode P3 (~ 34 cm$^{-1}$) at low temperature ($\leq$ 150 K). We attributed the observed anomalous trend in frequency and linewidth of this low frequency phonon to the electron-phonon coupling. Furthermore, our work also highlights the temperature induced tuning of resonance condition via understanding the intensity of phonon modes as a function of temperature.

[#]E-mail: deepu7727@gmail.com
[*]E-mail: pkumar@iitmandi.ac.in

## 1. Introduction

Reduced dielectric screening effect and large binding energy between electron and hole in atomically thin transition metal dichalcogenides (TMDCs), with common formula $MX_2$ (M= Mo, W and X= S, Se), gives rise to a large variety of excitonic transition energy states such as excitons and charged excitons. $A$ and $B$ excitonic energy states, commonly known as $A$ and $B$ excitons, respectively, are two very important excitonic features in these materials and appears due to direct excitonic transition from the minima of the conduction band to the maxima of split valence band at the $K$ point of the Brillouin zone [1-2]. These excitonic energy states play an important role in the Raman scattering process. In these materials, scattering cross-section of the phonon modes is strongly affected by laser excitation energy, which led to the modulation of Raman spectrum as a function of laser excitation energy [3-5]. For the case, when the incident excitation energy is far away from the excitonic transition energies states, only the Brillouin zone centre optical phonons ($q \sim 0$, where $q$ is the momentum of phonons) participates in the Raman scattering. On the other hand, when the incident excitation energy is close to one of the excitonic energy states, it results into a resonance effect. The resonance effect not only enhances the Raman scattering intensity of phonons but also the phonons from the entire Brillouin zone participates in the Raman scattering, which led to the observation of first order acoustic, second and higher order phonons from the high symmetry such as $M$ and $K$ points of the Brillouin zone. Moreover, the resonance effect may also break the selection rule which allows the observation of backscattering forbidden Raman and infrared active modes in the Raman scattering spectrum [6].

Electron-phonon coupling (EPC), a very important parameter in condensed matter physics, plays a key role in charge carrier mobility, heat generation and dissipation, phonon transport and phonon heat conduction and superconductivity [7-11]. In addition to the observation of multi-phonon modes in the Raman spectrum, resonance Raman scattering has also been proven to be an effective and powerful technique to investigate EPC in two-dimensional (2D) materials. The variations in the layer thickness, temperature, and energies of the incident laser excitation are the key parameters to investigate the EPC in these materials. *Bonini et al* determine the EPC via understanding the linewidth of phonons in graphite and graphene using first-principle calculation [12]; on the other hand, *Cong et al* evidence the EPC in graphene using temperature dependent Raman study of the low frequency phonon mode [13]. *Hinsche et al* investigated the EPC in monolayer $WS_2$ via understanding the spin splitting of the electronic band structure [14]. *Chakraborty et al* investigated EPC for the top-gated single layer $MoS_2$ using electron doping (voltage) dependent Raman study [15] and *Guo et al.* reported EPC in $MoS_2$ using ultrafast 2D visible/far-IR spectroscopy [16]. Recently, the signature of EPC has also been reported in TMDCs using resonance Raman scattering. *Rao et al* investigated the EPC in bulk and nano flakes of $MoS_2$ under the resonance effect [17]. *Paul et al* have also reported the signature of EPC via understanding the temperature dependent frequency and linewidth of the phonons in $MoS_2$ and $MoTe_2$ under the resonance effect [18]. Although a lot of efforts have been devoted previously to understand the EPC in TMDCs, but full understanding of EPC in TMDCs is still lacking.

In this work, we conducted the temperature dependent Raman measurements, in a temperature range of 4 to 330 K and focusing on low-frequency range of 10-300 cm$^{-1}$, on both three layers (3L) and ~ five-six layers adopted as few layers (FL for short) $MoS_2$. In order to achieve the resonance condition, we excite the spectra using 633 nm laser (1.96 eV) as this is very close to the energy of *B* exciton at room temperature, resulting in resonance effect. To the best of our knowledge temperature dependent Raman study on layered $MoS_2$ under the resonance effect for extreme low frequency range (≤ 100 cm$^{-1}$) as well as in very low temperature regime (≤ 80 K) has not been reported so far. Temperature dependent behaviour of multi-phonon modes have been crucial to understand the underlying physics associated with layered $MoS_2$. We investigated the EPC in addition to phonon-phonon coupling via understanding the temperature dependence of the frequency and linewidth of the phonon modes in the very low frequency region.

## 2. Results and Discussions

### 2.1 Low frequency phonons in MoS$_2$ under the resonance effect

Low frequency Raman spectrum of FL and 3L MoS$_2$, in the spectral range of 10-70 cm$^{-1}$ recorded at room temperature, is illustrated in Fig. 1 (a) and 1(b), respectively. Spectra are fitted with a sum of Lorentzian function to extract the self-energy parameters such as peak frequency ($\omega$), linewidth ($\gamma$) and intensity of the phonon modes. For the case of FL MoS$_2$, we observed two very weak peaks P1 and P2 located at ~ 14.9 and 26.5 cm$^{-1}$, respectively. The observed mode's frequency of P1 and P2 is very close to the frequency of interlayer shear or breathing modes in MoS$_2$ [19]. For 3L, in contrast to FL MoS$_2$, we could not observe any clear signature of P1 and P2 modes may be because of the very weak signal. Further, a broad and strong feature P3 centred at ~ 34 cm$^{-1}$ is observed in both FL and 3L MoS$_2$ which indicates the thickness-independent nature of this peak in line with the previous reports [19-20]. We also observed that peak P3 is slightly broader in case of 3L than that in FL MoS$_2$. In the literature, peak P3 is reported to be visible only if the spectrum is under the excitation with the laser excitation energy of 1.96 eV which is close to the $B$ (~1.95 eV at 300 K) excitonic transition states at $K$ point. Generally, below 100 cm$^{-1}$, the interlayer shear and breathing modes, which appear due to the interlayer interactions between the adjacent MoS$_2$ single layers, have been reported in layered MoS$_2$ and other 2D materials. Therefore, both interlayer shear and breathing modes are expected to be strongly dependent on the number of layers. However, we found that the peak position of the P3 mode is independent on the number of layers, suggesting it could not be assigned as interlayer shear or breathing mode. Zeng *et al* [20] attributed its origin to the electronic Raman scattering associated with the splitting in a conduction band at K points of the Brillouin zone due to the spin-orbit coupling. While another work carried out by *Lee et al* [19] contradicts its origin due to splitting of conduction bands. In order to understand the symmetry and polarization dependent of the P3, we performed polarization dependent measurements for both FL and 3 L MoS$_2$. The polarization dependent measurements were done by rotating the direction of the scattered light with an angle ($\theta$) by keeping fixed the position of the sample and direction of the incident light. Figure 1(c) and 1(d) show the room temperature Raman spectra in the spectral range of 10-75 cm$^{-1}$, collected at different rotation angles of scattered light with respect to fixed incident light. We observed that the intensity of the P3 mode is maximum at $0^0$ (i.e., when the incident and scattered light are parallel; parallel configuration), while the intensity becomes minimum at $90^0$ (i.e., when the incident and scattered light are perpendicular to each other; perpendicular configuration). Our polarisation dependent nature of mode P3 contradicts with the previous report where this mode is reported in both parallel and cross polarization configuration [20]. This observation indicates that its origin and symmetry assignment is still not clear and need further theoretical/experimental studies. Figure 1 (e) and 1 (f) demonstrate the Raman spectrum in the spectral range of 100-300 cm$^{-1}$, collected at 300 K, for both FL and 3L MoS$_2$, respectively. Within this spectral range, an intense mode P7 centred at ~177.6 cm$^{-1}$ and ~178.2 cm$^{-1}$ is observed for FL and 3 L MoS$_2$, respectively. In the literature, its (mode P7) origin has been attributed to the combination of Stokes and Anti-Stokes processes, where an optical ($A_{1g}$) phonon would be created (Stokes process) while an acoustical longitudinal ($LA$) phonon would be annihilated (Anti-Stokes process) at the $M$ symmetry point of the Brillouin zone and it may be assigned as $A_{1g}(M) - LA(M)$ [21]. Furthermore, we also observed a few very weak Raman scattering features (P4-P6 and P8-P11) at lower and higher frequency side of P7 mode. The appearance of these peaks in Raman spectrum may be attributed to the resonance effect and these observed features may be assigned as either first or second order Raman scattering modes from the $M$ or $K$ symmetry point of the Brillouin zone. For FL MoS$_2$, these modes are observed at ~123 cm$^{-1}$ (P4), ~134.5 cm$^{-1}$ ($E^1_{2g}(M)-LA(M)$; P5), ~148.5 cm$^{-1}$ ($TA(M)$; P6), ~186.3 cm$^{-1}$ ($TA(K)$; P8), ~208.2 cm$^{-1}$ (P9), ~227.9 cm$^{-1}$ ($LA(M)$; P10) and ~247.5 cm$^{-1}$ (P11). For the 3L MoS$_2$, these are observed at ~121.8 cm$^{-1}$ (P4), ~128.8 cm$^{-1}$ ($E^1_{2g}(M)-LA(M)$; P5), ~149.2 cm$^{-1}$ ($TA(M)$; P6), ~186.7 cm$^{-1}$ ($TA(K)$; P8), ~211.8 cm$^{-1}$ (P9), ~228.6 cm$^{-1}$ ($LA(M)$; P10). The symmetry assignment of the modes was done in accordance with the earlier studies on MoS$_2$ [22-23]. Figure 2 (a) and (b) show the temperature evolution of the Raman spectrum for FL and 3L MoS$_2$, respectively. Few interesting observations are given as: (i) Peak P1 is found to be intense at high temperature. (ii) P2 is well resolved from the P3 above ~ 180 K, while it merge into P3 below this temperature. (iii) Peak P4 is clearly visible only above ~80 K for the FL MoS$_2$, while it is visible only above ~ 200 K for the case of 3L MoS$_2$. (iv) For both FL and 3L MoS$_2$, Peak P8 is found to be narrow and well resolved from peak P7 in low temperature regime while it appears as shoulder of P7 at high temperature, see Fig. right panel of 2 (a) and 2(b). (v) Peak P7 is clearly visible up to our lowest recorded temperature (i.e. 4 K) for FL MoS$_2$, surprisingly it disappears completely below ~30 K for the case of 3L MoS$_2$. The appearance/disappearance of these modes with temperature may occur due to variations in resonance conditions which will be discussed in the next section in more detail.

**2.2. Effect of temperature on the intensity of the phonon modes**

In this section, we have focused on the intensity of the phonon modes as a function of temperature. In these 2D materials, intensity of the phonon modes and its dependence on the temperature as well as thickness of the materials may provide crucial information about the electronic and optical properties of the materials. The intensity of the phonons and its dependence on the temperature has been ignored in most of the previous literature. In order to understand quantitatively the intensity of the phonon modes as a function of temperature, we have extracted intensity of the individual phonon using the Lorentzian function. Figure 3 (a) and 3 (b) show the temperature dependent intensity of the few selected phonon modes for the case of FL MoS$_2$, temperature dependent evolution and fitting is shown in Fig.S1 and S2, see supplementary information. Following observation can be made: (i) For FL MoS$_2$, intensity of the P3 mode increases with decreasing temperature and become maximum at ~ 280 K. Below 280 K, a sharp decrease in intensity is observed and reaches minimum at ~100 K; on further cooling it again starts to increase till the lowest recorded temperature (i.e. 4 K). (ii) A sharp decrease in intensity of the mode P7 and P8 with decreasing temperature from ~300 to ~150 K is observed and below ~150 K intensity of the P7 (P8) mode decreases slightly (remains nearly constant). (iii) Intensity of the P9 and P10 mode is found to be maximum at highest and lowest recorded temperature and minimum at ~150 K, see inset in Fig 3(b). Figure 3 (c) and 3 (d) show the temperature dependent Raman intensity for the selected phonon modes for 3L MoS$_2$. Following observation can be made: (i) Intensity of the P3 mode decreases sharply from 330 to ~ 100 K, and an increase is observed on the further decrease in temperature below 100 K. (ii) Intensity of the P7 mode decreases with lowering the temperature from 330 to ~30 K, and below 30 K it vanished. The intensity of the P8 mode decreases sharply from 330 to ~100 K and below this temperature a slight increase is observed.

Now, we will focus on understanding the temperature dependent of the phonon's mode intensity. Within the semi-classical approximation, Raman scattering intensity of the phonon modes is given as $I_{int} = |\hat{e}_s^t . R . \hat{e}_i|^2$, where $\hat{e}_s^t$, R, $\hat{e}_i$ and $\hat{e}_s$ are the transpose of $\hat{e}_s$, Raman tensor and the unit vector of the incident and scattered light of polarization, respectively. This approximation has been proved to be best to describe the intensity of first order modes as a function of either change in direction of incident or scattered photon with known Raman tensor of the phonon modes [24-25]. However, it fails to explain the temperature dependent Raman scattering intensity of the phonon modes because it depends only on the Raman tensor and polarization direction of incident or scattered light. Furthermore, it also fails to explain the Raman scattering intensity of the second or higher order phonon modes. Under the non-resonant condition, the temperature dependent intensity of the first order, second order combination and difference phonon modes could be described using the Bose-Einstein (BE) function. For the case of Stokes Raman scattering process, the Raman intensity of a phonon (creation of a phonon; first order process), sum of two phonons (creation of two phonons; second order process) and difference of two phonons (creation of one phonon while annihilation of other one ; second order process) modes may be given as $(n_1+1)$, $(n_1+1)(n_2+1)$ and $(n_1+1)(n_2)$, respectively; where $n$ is the BE factor, is given as $n=[\exp(\hbar\omega_{ph}/k_bT)-1]^{-1}$ [26]. These expressions have been widely used to understand the temperature dependence of the intensity of the phonon modes under the non-resonant condition. However, these fail to explain the Raman scattering intensity of the phonon modes in the vicinity of resonance. Temperature dependent intensity of the phonon modes as shown in Fig. 3 is clearly reflecting that the BE expression is not sufficient to understand the intensity as a function of temperature. This is because the BE expression for intensity of the phonons depends only on the temperature, and therefore it doesn't care about the effect of the incident photon excitation energies on the intensity of the phonons. In order to fully understand the temperature dependent intensity, quantum mechanical picture needs to be invoked and is widely used to understand the temperature dependent intensity of the phonon in vicinity of the resonance effect.

It should be noted that $MoS_2$ and other TMDCs systems exhibit a large variety of excitonic energy states. Selectively choosing energies of the excitation laser source close to the energies of these transition's energy states, one may selectively modulate the Raman scattering intensity of the phonon modes. Moreover, the excitonic energy states are strongly dependent on temperature indicating that the crossover in resonance and non-resonance conditions may not be obtained only by varying the excitation laser energy but also could be achieved by varying the temperature. This crossover between resonance and non-resonance with variations in temperature may led to modulation in the intensity of the phonon modes. The intensity of the phonon in first order Raman scattering considering the quantum mechanical picture can be given as [26-27]

$$Int. = \left| \sum_{g,i,i'} \frac{<g|H_{op_2}|i'><i'|H_{el-ph}|i><i|H_{op_1}|g>}{(E_L - \Delta E_{ig})(E_s - \Delta E_{i'g})} \right|^2 \quad (1)$$

where, $|i>$, $|i'>$ and $|g>$ represent the intermediate states and the ground state, respectively. $H_{op_1}$ and $H_{op_2}$ are the Hamiltonians representing the electron-photon (absorption) and electron-photon (emission) process, respectively; and $H_{el-ph}$ is the electron-phonon interaction Hamiltonian. $\Delta E_{ig} = (E_i - E_g + i\gamma_1)$ and $\Delta E_{i'g} = (E_{i'} - E_g + i\gamma_2)$ represents the energy difference between the intermediate state $E_i$ and the ground states $E_g$ and intermediate states $E_{i'}$ and the final/ground states $E_g$, respectively. $\gamma_1$ and $\gamma_2$ are related to the finite lifetime of the intermediate states. $E_L = \hbar\omega_L$ and $E_S = \hbar\omega_S = \hbar\omega_L - \hbar\omega_{ph}$ correspond to the energy of incident and scattered photons, respectively. $\omega_{ph}$ corresponds to the frequency of the participating phonons. If the intermediate states are real excitonic energy states, then $E_i - E_g$ and $E_{i'} - E_g$ could represent to $A$ and $B$ the excitonic energy states in MoS$_2$ and other such materials. When incident laser excitation energy approaches one of these excitonic energy states then there will be the resonance effect. The numerator (matrix elements terms) may become weak compare to the denominator in the vicinity of the excitonic resonance effect and it may be considered as constant. In this case Raman scattering intensity of phonon may be written as [28]

$$Int. \propto \left| \frac{1}{[E_x(T) - E_L + i\gamma_x(T)][E_x(T) - E_s + i\gamma_x(T)]} \right|^2 \quad (2)$$

where $E_x(T)$ and $\gamma_x(T)$ are the temperature dependent transition energies and linewidth correspond to excitonic energy states ($x = A$, $B$ etc.), respectively. Above Eq$^n$ (2) suggests that resonance conditions could be changed either by changing the incident excitation laser energy or temperature. Generally, the energy of the excitonic states increases with lowering the temperature. The change in the excitonic energy with temperature results into the deviation in resonance condition which may be reflected in the Raman spectrum. In the present case, we have excited the Raman spectra with fixed incident photon energy which resonates with $B$ exciton at room temperature. For the case of 3L MoS$_2$, the $E_A$ increases from ~ 1.84 eV at 300 K to ~ 1.89 eV at 4 K while $E_B$ increase from ~ 1.95 eV at 300 K to ~ 2.03 eV at 4K with lowering the temperature [29]. This indicating that with decreasing temperature, $E_B$ is moving away from the $E_L$, while $E_A$ is approaching towards $E_L$ indicating the tuning of resonance condition from $B$ exciton at room temperature to $A$ exciton at the lowest recorded temperature i.e. 4 K. This is also confirmed by the intensity trend of P3 as it is predominantly visible only when the energy of the excitonic transition states is ~1.96 eV. For both FL and 3L MoS$_2$, we observed that intensity of the P3 mode decreases with decreasing temperature from 300 K and become minimum at ~ 100 K, may be because $E_B$ moves away from the 1.96 eV incident photon energy. While below 100 K its intensity starts to increase may be because $E_A$ moves towards 1.96 eV. This observation clearly reflecting the tuning of resonance condition from $B$ exciton at 300 K to the $A$ exciton at low temperature.

## 2.3. Electron-phonon coupling

Figure 4 (a and b) and 4 (d and e) show the temperature dependent frequency and linewidth of the P3 mode for FL and 3L MoS$_2$. For both systems, we observed the hardening and narrowing (normal temperature dependent behaviour) of the P3 mode with lowering the temperature from 330 to ~ 120-150 K. On the other hand, below ~150 K, an anomalous softening and broadening is observed. Furthermore, we notice that the decrease in frequency is significantly large in 3L compared to FL below ~120-150 K, while change in linewidth is nearly the same in both systems. The observed anomalous softening and broadening of P3 mode with decreasing temperature below ~ 150 K could not be explained only the considering anharmonicity effect (decaying of an optical phonons into acoustic phonons) because this effect led to the hardening and narrowing of phonon mode with decreasing temperature. Spin-phonon coupling, another possible parameter, which may also lead to the anomalous softening and broadenings [30-31]. But this effect could also be excluded from the fact that the $MoS_2$ is not a magnetic material. The previous reports show the involvement of EPC results into the anomalous temperature dependent trend in the linewidth of the phonon mode in these kinds of 2D materials [13, 32]. The observed anomalous behaviour of P3 below ~ 150 K suggest its origin the EPC. For FL $MoS_2$, $\Delta\omega_{P3}=(\omega(150\,K)-\omega(4\,K))$ and $\Delta\gamma_{P3}=(\gamma(150\,K)-\gamma(4\,K))$ are found to be ~0.6 cm$^{-1}$ and ~5.3 cm$^{-1}$, respectively. While for the case of 3L $MoS_2$, $\Delta\omega_{P3}=(\omega(150\,K)-\omega(4\,K))$ and $\Delta\gamma_{P3}=(\gamma(150\,K)-\gamma(4\,K))$ are found to be ~4.6 cm$^{-1}$ and ~5.7 cm$^{-1}$, respectively. *Chakraborty et al* reported that the increase in the electron concentration leads to the softening and broadening of the Raman active $A_{1g}$ phonon mode by 4 cm$^{-1}$ and 6 cm$^{-1}$, respectively [15]. Observed change in the frequency and linewidth (except in frequency for FL $MoS_2$) in temperature range of 4 to 150 K is very close to the changes reported by *Chakraborty et al* suggesting that the softening and broadening of mode P3 with lowering temperature below 150 K may be due to the contributions of EPC, which was also confirmed by a theoretical model consisting contributions from both EPC and phonon-phonon coupling (PPC).

Normally, the phonon mode's frequency decreases with increasing temperature due to the lattice anharmonicity and thermal expansion. A well-defined model has been proposed by the scientific community to understand the role of individual contributions to the shift in frequency of the phonon mode with temperature. Since the various factors are affecting the frequency of the phonon modes as a function of temperature, therefore the unambiguous estimation of individual role of EPC from the phonon mode's frequency is very challenging. On the other hand, the linewidth of the phonons could be the best parameter to extract the EPC. Therefore, we have only focused on the linewidth of phonons, especially mode P3 in order to extract the EPC. Generally, the linewidth of the phonon mode as a function of temperature is predominantly affected by PPC and it increases with increasing temperature. This may be understood as: the population of phonons is governed by the BE distribution function ($n = [\exp(\hbar\omega_{ph}/k_b T) - 1]^{-1}$) which is expected to increase with increasing temperature. The increase in population of the phonons with increasing temperature lead to the decrease in phonon's lifetime. It should be noted that linewidth of the phonons is inversely proportional to lifetime of the phonons ($linewidth \propto 1/\tau$, where $\tau$ is the lifetime of phonons). Therefore, the increase in linewidth of the phonons is expected with the rise in temperature. The observed trend in linewidth of the P3 mode above 150 K may be understood considering the PPC. However, the observed anomalous trend of P3 below 150 K could not be explained using PPC and it may be understood considering the EPC contribution. Temperature dependent linewidth of the phonons considering the contributions from both EPC and PPC may be given as [12]

$$\gamma(T) = \gamma_{ph\text{-}ph}(T) + \gamma_{e\text{-}ph}(T) \tag{3}$$

Where first term $\gamma_{ph\text{-}ph}(T)$ corresponds to the contribution from PPC and the second term $\gamma_{e\text{-}ph}(T)$ represents the contributions from the EPC. At finite temperature, the contributions from the three-phonon anharmonic effect to the linewidth of phonons as suggested by Klemens can be given as [33] $\gamma_{ph-ph}(T) = \gamma_{ph-ph}(0) + C(1 + \frac{2}{e^x - 1})$, where $x = \hbar\omega_0/2k_B T$. C is a self-energy constant fitting parameter corresponding to the decaying of an optical phonon into two acoustic phonons with equal energy but opposite momentum and $\gamma_{ph-ph}(0)$ is the linewidth at 0 K. The contributions from the EPC to the linewidth of phonons may be given as [12] $\gamma_{e-ph}(T) = \gamma_{e-ph}(0)[(\frac{1}{e^{-y}+1}) - (\frac{1}{e^y+1})]$, where $y = \hbar\omega_0/2k_B T$ and $\gamma_{e-ph}(0)$ is the linewidth at 0 K. The terms $\gamma_{ph-ph}(T)$ [$\gamma_{e-ph}(T)$] is expected to increase [decrease] with increasing temperature. The solid red lines in Fig. 4 (b) and 4(e) are the fitting curves using above Eq$^n$ (3) and best fit values of parameters $C$, $\gamma_{ph-ph}(0)$ and $\gamma_{e-ph}(0)$ are found to be ~0.06, ~7.3 cm$^{-1}$ and ~10.9 cm$^{-1}$, respectively for FL; while these parameters are found to be ~0.2, ~4.1 cm$^{-1}$ and ~13.0 cm$^{-1}$, respectively for 3L MoS$_2$. In order to understand the individual contributions from EPC and PPC to the linewidth of the P3, we have separately plotted the $\gamma_{e\text{-}ph}(T)$ and $\gamma_{ph\text{-}ph}(T)$ as a function of temperature. Figure 4 (c) and 4 (f) show the individual contributions to the linewidth of P3 mode from PPC (blue) and EPC (red) for FL and 3L MoS$_2$, respectively. For both FL and 3L MoS$_2$, the function $\gamma_{e\text{-}ph}(T)$ increases slightly with cooling from our highest recorded temperature (330 K) to ~ 150 K, and below ~150 K we observed a sharp increase till the lowest recorded temperature (4 K), clearly suggesting that the broadening of P3 mode below 150 K is indeed originate due to the strong contributions of EPC.

**Conclusion**

In conclusion, herein we have presented our study devoted to the observation of signature of strong electron-phonon coupling in layered MoS$_2$. Our resonance Raman scattering study provides clear spectroscopic evidence of electron-phonon coupling in addition to phonon-phonon coupling and it dominates over phonon-phonon coupling in the low temperature regime. Further we also observed that the intensity of the phonon modes is strongly modulated as a function of temperature reflecting the modulation of resonance conditions with temperature for both 3L as well as FL MoS$_2$.

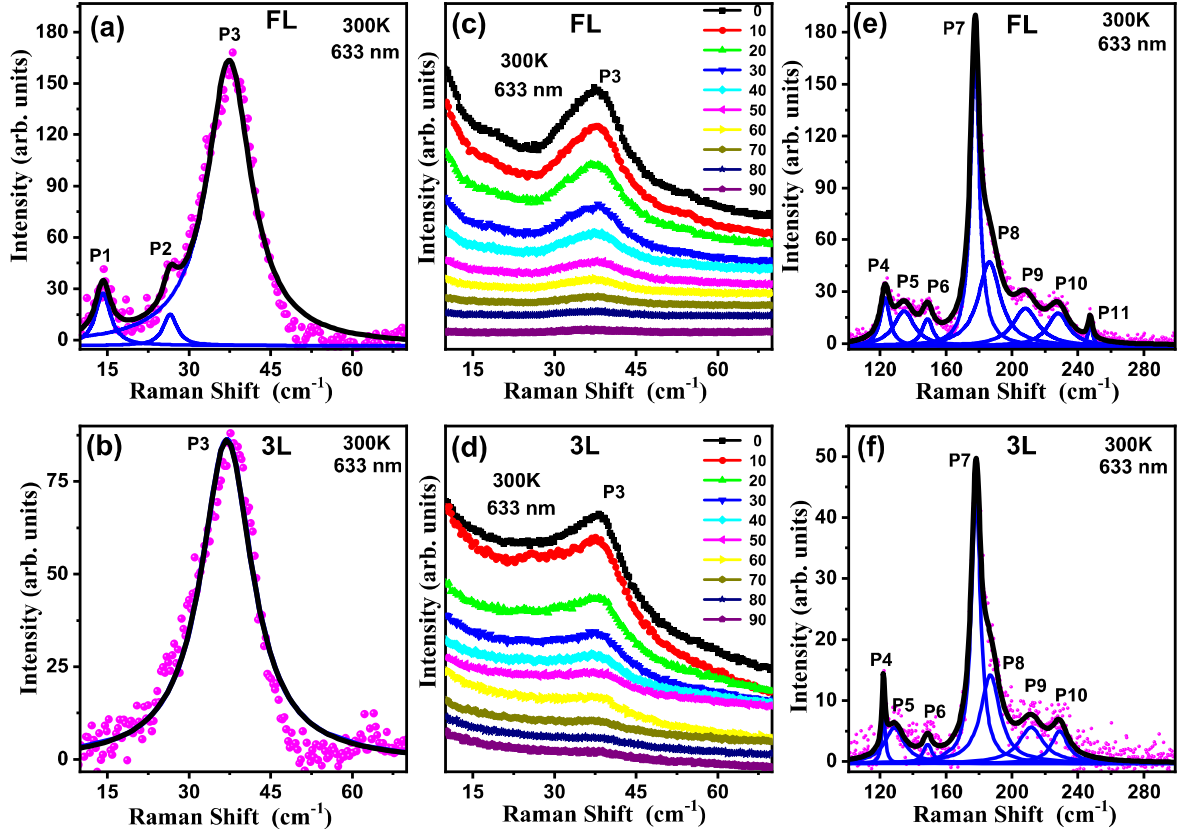

**Figure 1:** (a) and (b) Room temperature Raman spectra in a frequency range of 10-70 cm$^{-1}$ for the FL and 3L MoS$_2$, respectively. (c) and (d) Room temperature Raman spectra in frequency range of 10-70 cm$^{-1}$ recorded at different polarization angles from $0^0$ to $90^0$ for the FL and 3L MoS$_2$, respectively. (e) and (f) Room temperature Raman spectra in a frequency range of 100-300 cm$^{-1}$ for the FL and 3L MoS$_2$, respectively.

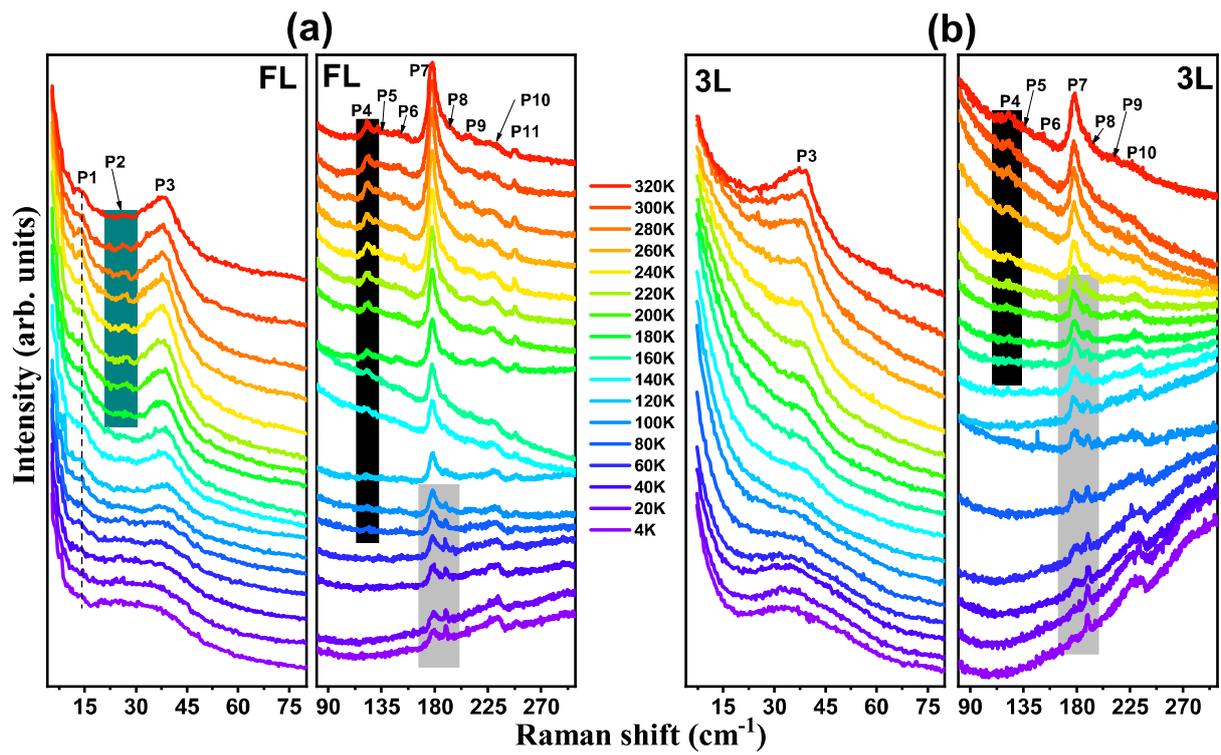

**Figure 2:** (a) and (b) show the temperature evolution of the Raman spectrum for FL and 3L MoS$_2$, respectively.

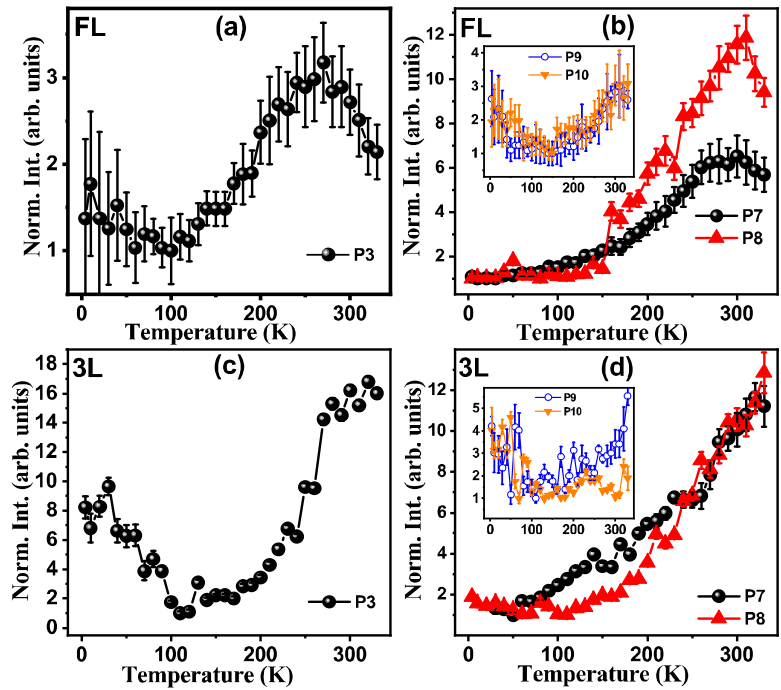

**Figure 3:** (a) and (b) Temperature dependence of the normalized intensity of P3, P7 and P8 phonon modes for FL $MoS_2$. (c) and (d) Temperature dependence of the normalized intensity of P3, P7 and P8 phonon modes for 3L $MoS_2$. Insets in panel (b) and (d) are the temperature dependence of the normalized intensity of P9 and P10 phonon modes.

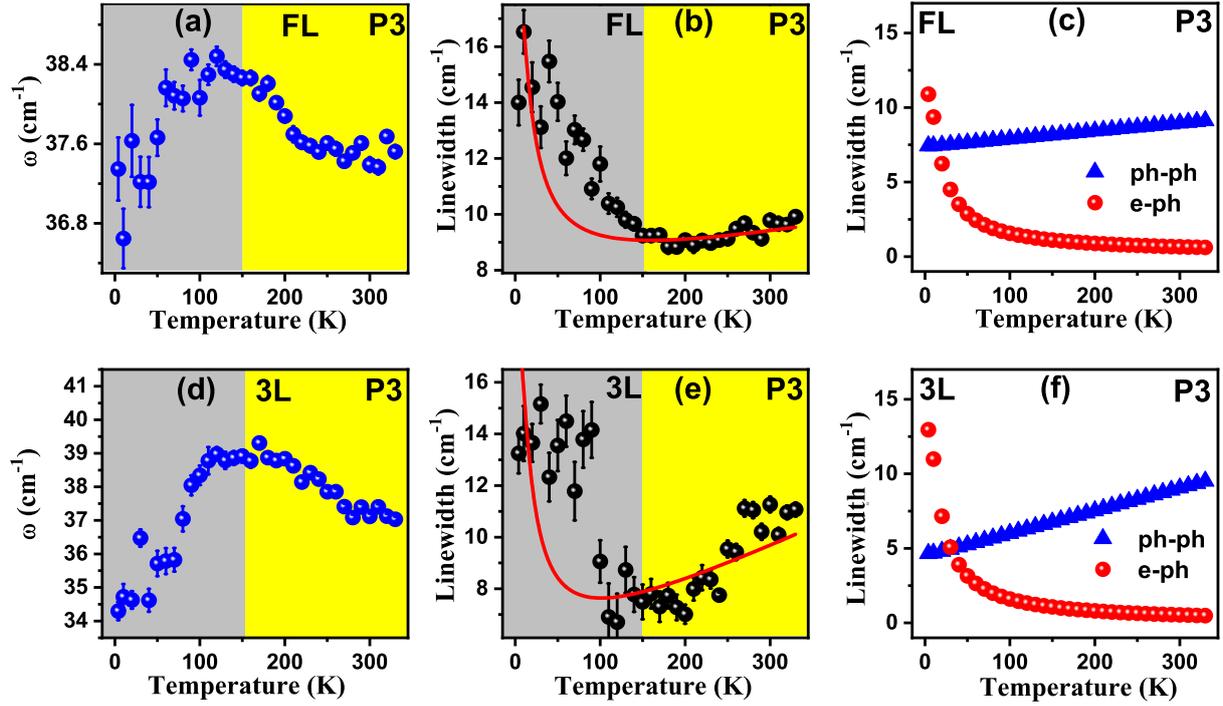

**Figures 4:** (a) and (d) Temperature dependent frequency ($\omega$) of the P3 phonon modes for FL and 3L MoS2, respectively. (b) and (e) Temperature dependent linewidth ($\gamma$) of the P3 phonon modes for FL and 3L MoS2, respectively. Yellow and grey shaded area demonstrate the contributions from PPC and EPC coupling to the frequency and linewidth of the P3 mode, respectively. (c) and (f) show the individual conurbations from PPC and EPC to the frequency and linewidth of the P3 mode for FL and 3L MoS2, respectively.

**Supplementary Information**

# Tunable Resonance and Electron-Phonon Coupling in Layered $MoS_2$

Deepu Kumar[1#], Nasaru Khan[1], Rahul Kumar[2], Mahesh Kumar[2], Pradeep Kumar[1*]

[1]*School of Physical Sciences, Indian Institute of Technology Mandi, 175005, India*
[2]*Department of Electrical Engineering, Indian Institute of Technology Jodhpur, 342037, India*

[#]E-mail: deepu7727@gmail.com

[*]E-mail: pkumar@iitmandi.ac.in

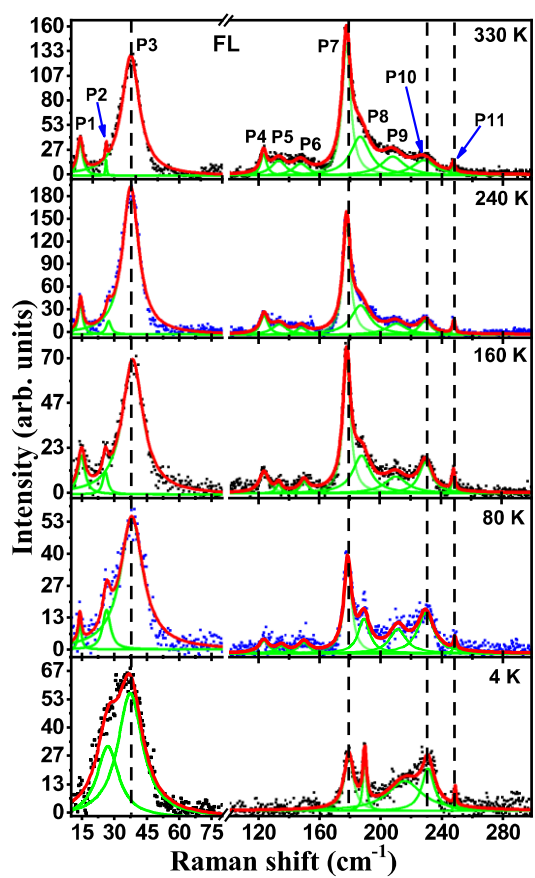

**Figure S1:** Raman spectra for FL $MoS_2$ recoded at different temperature 4 K, 80 K, 160 K, 240 K and 330 K. The solid red lines are Lorentzian fits to the total spectrum, and the green lines are the Lorentzian fit to the individual peak.

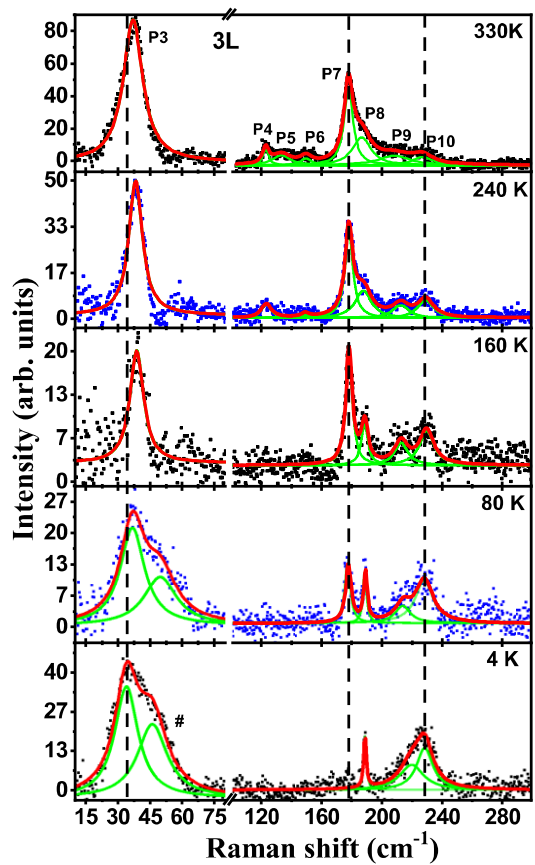

**Figure S2:** Raman spectra for 3L $MoS_2$ recoded at different temperature 4 K, 80 K, 160 K, 240 K and 330 K. The solid red lines are Lorentzian fits to the total spectrum, and the green lines are the Lorentzian fit to the individual peak. A new mode (marked by #) appear below ~100 K.

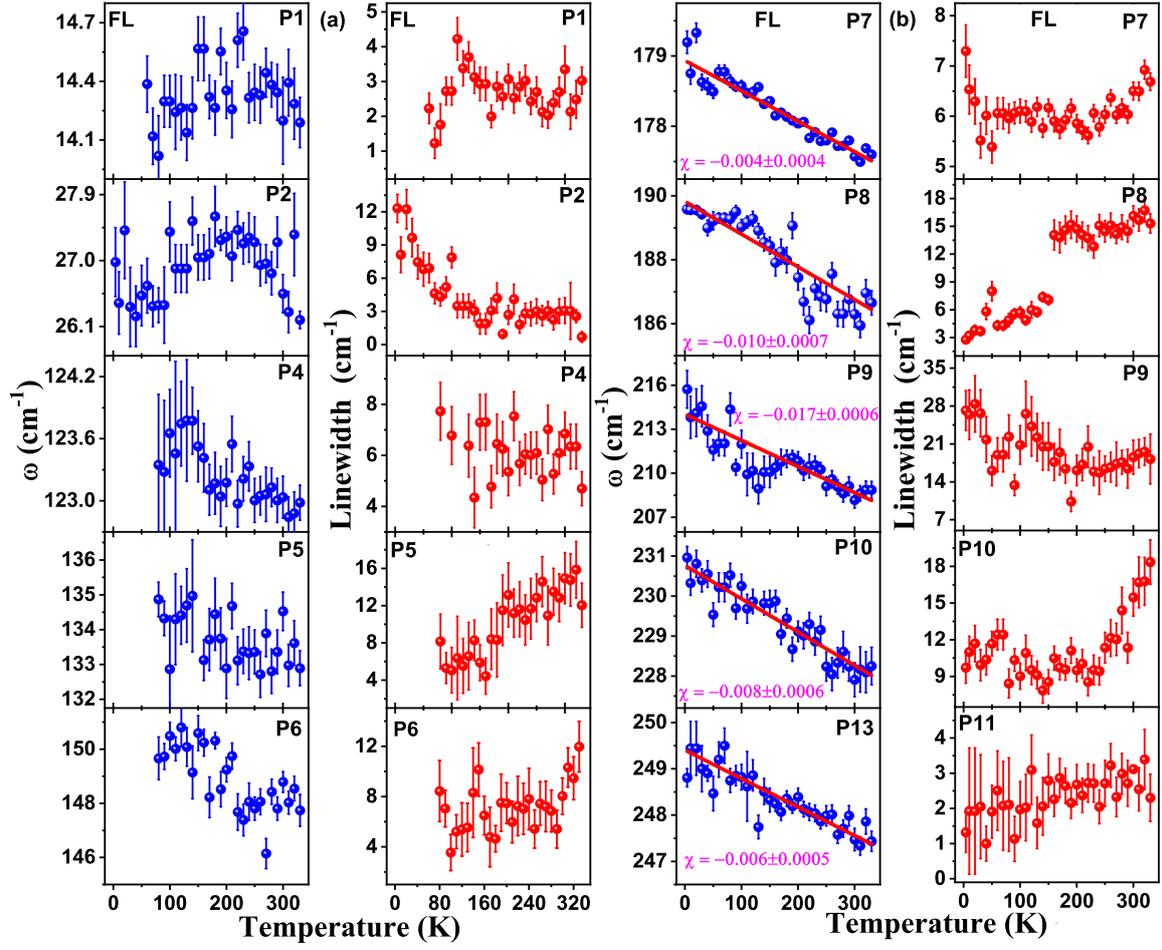

**Figure S3:** (a) and (b) Temperature dependent frequencies and linewidths of the phonon modes for FL MoS$_2$. Red solid lines are the fitted curves using the linear approximation for temperature dependent frequency and is given as $\omega(T) = \omega_0 + \chi T$, where $\omega_0$ is the frequency at 0K and $\chi$ is s the first order temperature coefficient.

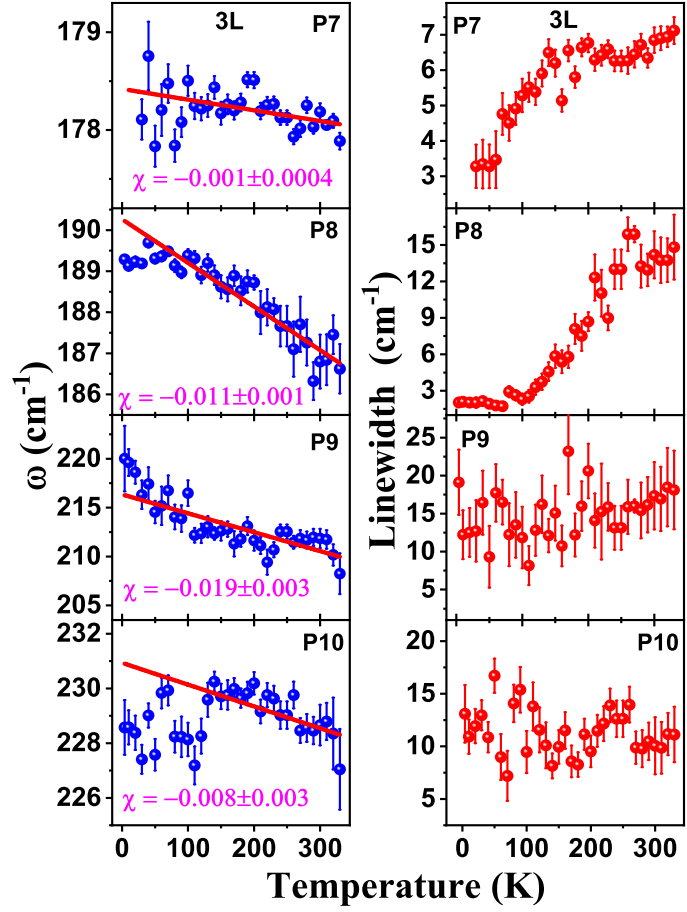

**Figure S4:** (a) and (b) Temperature dependent frequencies and FWHMs of the phonon modes for 3L MoS$_2$. Red solid lines are the fitted curves using the linear approximation for temperature dependent frequency and is given as $\omega(T) = \omega_0 + \chi T$, where $\omega_0$ is the frequency at 0K and $\chi$ is s the first order temperature coefficient.